\documentclass[letterpaper]{article}

\usepackage[T1]{fontenc}

\usepackage{geometry}
\usepackage{setspace}
\usepackage{hyperref}
\usepackage[style = chem-acs, articletitle, hyperref,
            maxnames=10,]{biblatex}
\usepackage{graphicx}
\usepackage{float}
\newfloat{scheme}{htbp}{los}
\floatname{scheme}{Scheme}
\floatname{chart}{Chart}
\newfloat{graph}{htbp}{loh}

\usepackage{chemformula} 
\usepackage[version = 4]{mhchem} 

\usepackage{authblk}

\author[1]{Robbie G. Hunt}
\author[1]{Gunnar K. Pálsson}
\author[1]{Matías P. Grassi}
\author[1,2]{Victoria Kabanova}
\author[1,2]{Alexey Vorobiev}
\author[1,*]{Gabriella Andersson}

\affil[1]{Department of Physics and Astronomy, Uppsala University, Box 516, SE-75120, Uppsala, Sweden}
\affil[2]{Institut Laue-Langevin, Grenoble, France}

\title{Probing deuterium-induced magnetic phase transitions in TbCo alloys with \textit{in-situ} polarized neutron reflectometry}

\date{*Email: gabriella.andersson@physics.uu.se}

\begin{document}

\maketitle

\begin{abstract}
 Hydrogen-based magneto-ionics is a promising approach for rapid magnetoelectric control of spintronic devices. Most investigations so far into the magneto-ionic manipulation of rare-earth transition-metal alloys have used electrochemical methods for evaluating the magnetoelectric properties, but this technique makes it difficult to discriminate between the effects of competing ionic species. In this work, we use atmospheric loading to evaluate the effect of an isotope of hydrogen, deuterium, on the magnetic properties of TbCo films using \textit{in-situ} polarized neutron reflectometry. With this approach, we are able to simultaneously measure the magnetization, thickness expansion and deuterium concentration of TbCo films. We quantitatively observe the deuterium concentrations at which the paramagnetic phase transition occurs for a Tb-rich film, and the weakening of out-of-plane magnetic anisotropy for a Co-rich film. For the Tb-rich film the expansion of the film thickness is the primary mechanism identified for the paramagnetic phase transition, while for the Co-rich film no thickness expansion is observed. We also find that an oxidized interface is insensitive to deuterium loading, but remains exchange coupled to the rest of the film and can be indirectly manipulated by loading of deuterium in the alloy. We expect these results to be directly translatable to that of hydrogen.
\end{abstract}

\section*{Keywords}

Magneto-ionics, ferrimagnetism, rare-earth transition-metal alloys, polarized neutron reflectometry.

\section{Introduction}

Magnetic materials have become integral to the design of computational devices and data storage, from magnetic-tape data storage \cite{dee2008magnetic} to giant magnetoresistive \cite{baibich1988GMR} and magnetic tunnel junctions \cite{bowen2001large} based devices. At the heart of their modern applications is the plethora of magneto-transport properties that allow for extremely precise read-out and manipulation of the magnetic state of a nanoscale magnetic layer. A significant challenge with implementing devices based on these properties into computational devices is the problem of Joule heating - if a magnetic field has to be applied to manipulate them then this must be generated using a current, which carries significant power losses that are dissipated as heat. To overcome this, there has been research into magnetoelectric materials and mechanisms, where an electric field can be used to manipulate the magnetic state directly, bypassing the need for a magnetic field by using, for example, multiferroic materials \cite{Mostovoy2024, ederer2005weak}, dilute magnetic semiconductors \cite{telegrin2022}, and voltage-controlled strain transfer \cite{hunt2025control, shuai2022local, franke2015reversible}. 

In recent years, magneto-ionics has become an increasingly popular method of manipulating magnetic properties via electric fields. The technique is extremely versatile and allows for a wide combination of materials and ionic species to manipulate magnetic properties directly through modification of the stoichiometry and crystal structure \cite{de2022voltage}. So far, it has been shown to have a wide scope in applicability with tuneable properties ranging from bulk-like, such as the saturation magnetization \cite{kutuzau2025additive, lopez2024room} and Curie temperature of a material \cite{yan2015electrical, zamani2013}, to interfacial such as the RKKY interaction \cite{kossak2023voltage, bauer2015magneto, ma2025magneto} or interface-induced perpendicular magnetic anisotropy \cite{bauer2015magneto}.

As a choice of ion, hydrogen has the advantage of having a well-understood atmospheric loading mechanism\cite{alefeld1978hydrogen1, alefeld1978hydrogen2} that allows the magneto-ionic tuning of a material to be studied independently of the presence of any additional ions. A completely ion-specific study can be difficult to achieve in electrochemical systems as there can be deleterious ions (such as O$^{2-}$) which also migrate in response to an electric field, and so this atmospheric mechanism allows for a straightforward analysis of the hydrogen-specific changes. In exothermic metal-hydrogen systems, ionic hydrogen can easily diffuse into the metal and form metal hydride phases and this is a pre-requisite for there to be a strong magneto-ionic effect with hydrogen. As such, previous work looking at the effect of atmospheric hydrogen loading on magnetic properties have focused on combinations of materials where at least one element has a high affinity for hydrogen: Pd in CoPd alloys \cite{das2018detection}, nanocomposites \cite{gossler2021nanoporous} or Co/Pd multilayers \cite{bischoff2024magneto}, Nb in Fe/Nb multilayers \cite{klose1997continuous}, V in Fe/V superlattices \cite{labergerie2001hydrogen, hjorvarsson1997reversible}, to name a few examples. 

In this work we investigate the magneto-ionic effect of deuterium in amorphous TbCo alloys through atmospheric loading. By choosing to use polarized neutron reflectometry (PNR) we obtain the possibility to measure the hydrogen concentration, magnetization profile, and changes in thickness simultaneously. For this experiment, we use deuterium (D) instead of hydrogen (H) due to the strong isotope effect on the neutron nuclear bound coherent scattering length, $b$. Deuterium ($^{2}H$) has a much larger absolute value of $b$ than hydrogen ($^1$H) , and this will considerably increase the sensitivity to small changes in the amount of loaded deuterium \cite{cousin2020introduction}. While there will be some differences in, for instance, the loading isotherms and diffusion constants of each isotope we expect that the results will be broadly transferable between D and H due to the similarity in ionic radii and electronic structures of both isotopes, as well as identical thickness expansion coefficients \cite{alefeld1978hydrogen1}. 

These rare-earth transition-metal alloys are ferrimagnetic, with the magnetic moments on the Tb and Co "sublattices" coupled antiferromagnetically and aligned antiparallel but with a non-zero net magnetization due to the difference in size between $m_{\mathrm{Tb}}$ and $m_{\mathrm{Co}}$. Depending on temperature and alloy composition, the larger moment can be on either the Tb or Co sublattice, i.e., the net magnetization is either Tb-dominated or Co-dominated, with a crossover at the compensation point where the net magnetization vanishes. The PNR measurements are sensitive to the net magnetization, not to the individual sublattice moments.

We study two phase transitions for magneto-ionic tuning: an in-plane to paramagnetic phase transition, and an out-of-plane to in-plane spin reorientation, both of which could be useful for device applications. To that end, we consider two alloy compositions: Tb$_{35}$Co$_{65}$, which has Tb-dominated in-plane magnetization, and Tb$_{14}$Co$_{86}$, which has a Co-dominated out-of-plane net magnetization prior to any ion loading. 


\section{Experimental Methods}

\subsection{Sample Growth}

Samples are grown via DC magnetron sputtering in a vacuum system with base pressure of 8$\times10^{-10}$ mbar or better. Prior to deposition substrates are degassed at 200 \textdegree C for 40 minutes to clean the substrate surface. Thin films are deposited with an Ar pressure of 3.3 mbar (purity 99.995\% and further purified using an inline gas filter) onto silicon substrates with a native oxide layer. The Tb$_x$Co$_{(100-x)}$ film is deposited directly onto the substrate using a co-sputtering technique with the Co power adjusted to obtain the desired composition. The sample is then transferred to an adjacent chamber with a base pressure of 5$\times10^{-8}$ mbar and a Pd layer is deposited with an Ar pressure of 3.4 mbar using industrial-grade Ar.

Samples are initially characterized by X-ray reflectometry (XRR) and Rutherford back-scattering spectrometry (RBS) to confirm the initial structure and Tb-Co composition, and magneto-optic Kerr effect (MOKE) measurements to verify the magnetic properties. XRR data is fit using GenX \cite{bjorck2007genx} and RBS data is fit using SIMNRA \cite{mayer1999simnra}. Fitting the XRR data confirms that the as-deposited structure of the two samples is Si/SiOx(2)/Tb$_{35}$Co$_{65}$(56)/Pd(8) and Si/SiOx(2)/Tb$_{14}$Co$_{86}$(68)/Pd(8), where all thicknesses are in nm. The thicknesses of both layers are chosen to optimize neutron reflectivity contrast between the TbCo and Pd layers, and to have enough features available to assist in fitting of neutron reflectivity data. The initial XRR fitting is used to determine the starting parameters for the polarized neutron reflectivity (PNR) fitting.

\begin{figure}
    \centering
    \includegraphics[width=\linewidth]{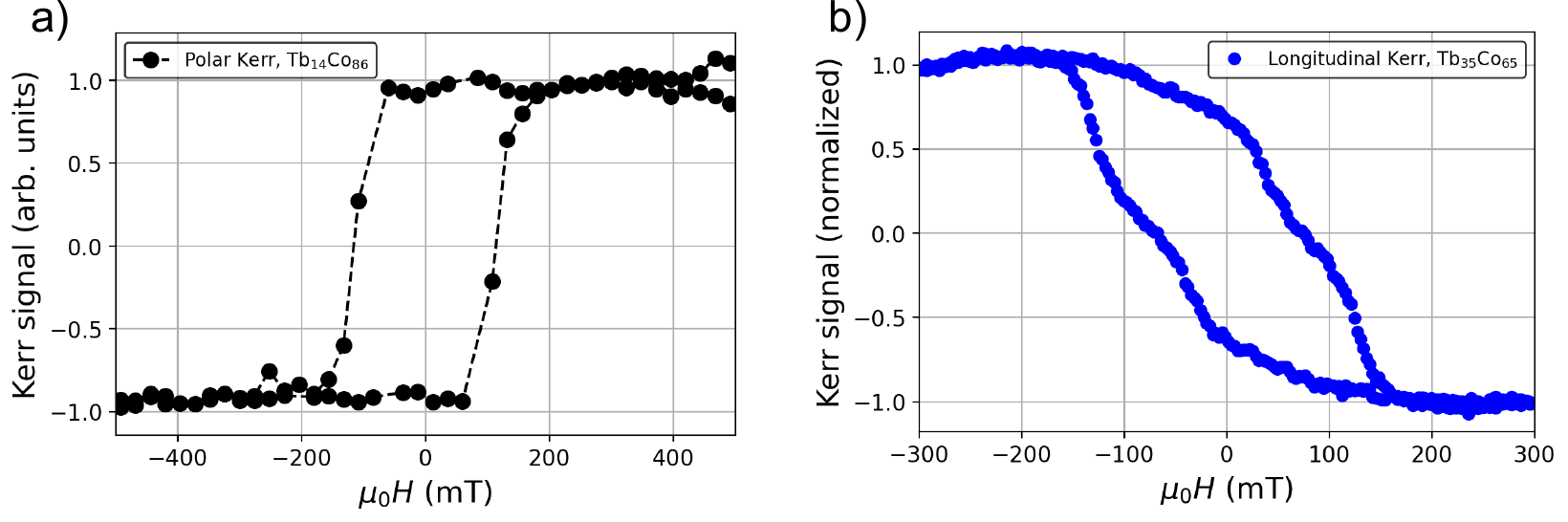}
    \caption{Normalized hysteresis loops for samples a) Tb$_{14}$Co$_{86}$ measured in a polar geometry and b) Tb$_{35}$Co$_{65}$ measured in a longitudinal geometry. Dashed lines are intended as a guide to the eye.}
    \label{fig:hysteresis}
\end{figure}

Hysteresis loops for as-deposited samples are shown in Fig. \ref{fig:hysteresis}. The Tb$_{14}$Co$_{86}$ sample ("Co-rich") exhibits a strong out-of-plane anisotropy with a square hysteresis loop in polar geometry, while the Tb$_{35}$Co$_{65}$ sample ("Tb-rich") has in-plane anisotropy demonstrated by the strong hysteresis loop measured in the longitudinal geometry. The magnetic behaviour exhibited is in line with previously reported properties for similar films \cite{ciuciulkaite2020magnetic, hunt2025control}. 


\subsection{Polarized Neutron Reflectometry}

Polarized neutron reflectometry is performed using the SuperADAM \cite{devishvili2013superadam, vorobiev2015recent} instrument at the Institut Laue-Langevin. Neutrons are polarized using a polarizing supermirror and spin-flipper to select the polarization of the neutrons prior to arriving at the sample. The neutron beam is monochromatic with a wavelength of $\lambda = 5.21$ Å. The entire pattern is measured on a 2D detector and an in-house software is used for reduction and overillumination correction of the data. The reduced data is then modelled and fit using GenX. The sample is mounted in a cryostat with a small heater on the back of the substrate, and connected to a turbo molecular pump and a deuterium bottle. Samples are allowed to outgas overnight and a base pressure of $5\times10^{-5}$ mbar is obtained.


The sample is initially held under vacuum at 320 K for the measurements of the unloaded sample, and an atmosphere of deuterium is then introduced using a D$_2$ gas bottle with 99.99\% purity. A hydrogen-sensitive pressure gauge is used to obtain exact pressures. The sample is left for at least one hour to reach a state of relative equilibrium before full scans are performed.

During this waiting time, measurements are continually taken in the $Q$-range of 0-0.04 \AA$^{-1}$ to track the loading process. This range of $Q$ encompasses the critical edge of the reflectivity pattern, which is extremely sensitive to changes in the nuclear scattering length density (SLD), allowing for a quick feedback loop of the loading process.

All measurements are performed with an in-plane magnetic field of 300 mT. This is sufficient to saturate the Tb$_{35}$Co$_{65}$ sample (see Fig. \ref{fig:hysteresis}), and so allows us to track the absolute change in the net in-plane magnetic moment from the difference in intensities of the two neutron polarizations, i.e. the spin asymmetry which is defined as $SA =\frac{R^+-R^-}{R^++R^-}$. 

For the Tb$_{14}$Co$_{86}$ sample the field is not sufficient to fully overcome the strong perpendicular magnetic anisotropy (PMA) and so instead we measure only a small rotation of the net moment in-plane, which is still significant enough to measure some degree of spin asymmetry. Any weakening of the PMA strength will then result in a greater reorientation of the in-plane magnetization component.

\section{Results and discussion}

\subsection{Magneto-ionic effect in in-plane magnetized alloys}

\begin{figure}
    \centering
    \includegraphics[width=\linewidth]{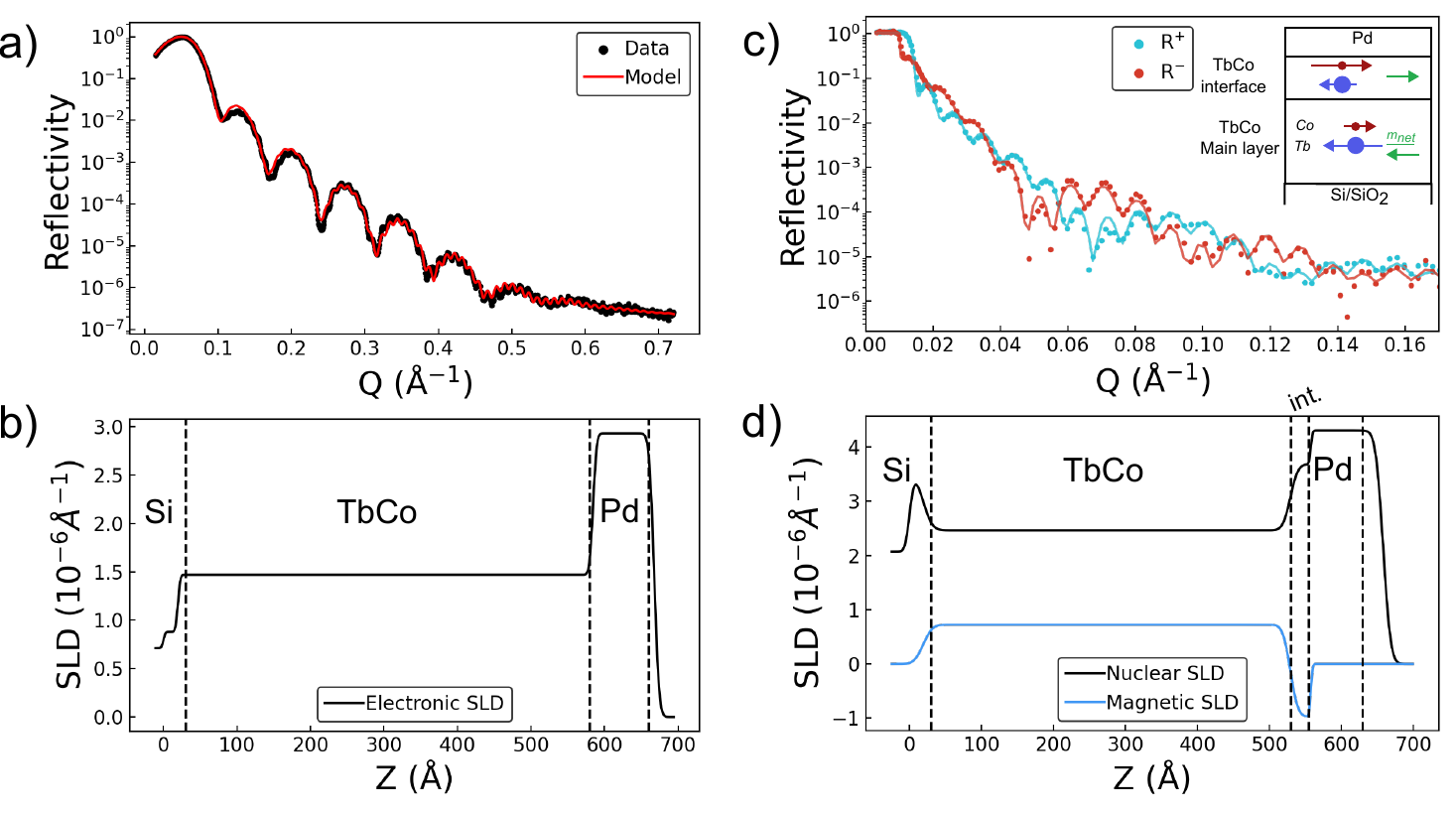}
    \caption{a, b) X-ray reflectivity data and electronic SLD profile model used to fit the reflectivity for the Tb$_{35}$Co$_{65}$ sample. c, d) corresponding data and model for the same sample as measured by PNR. Inset in c) is a schematic representation of the sample layer stack for the PNR model. Between the Pd and TbCo layer there is a small interfacial layer which has a  Co-dominated magnetization that is coupled antiparallel to $m_{\mathrm{net}}$ in the main TbCo layer.}
    \label{fig:tb-rich_model}
\end{figure}

We begin by discussing the results of the Tb$_{35}$Co$_{65}$ sample. To reiterate, this sample has in-plane magnetic anisotropy and is Tb-dominated. The initial model for the sample is shown in Fig. \ref{fig:tb-rich_model}b). From the XRR data alone the sample can be sufficiently described by a slab model of Si/TbCo/Pd. However in the PNR data several features emerge that necessitate a more complex model. The exact evolution of the model complexity is shown in the supplementary information. The final model we arrive at for the PNR data is Si/SiO2(1.5)/Tb$_{35}$Co$_{65}$(51)/TbCoOx(3)/Pd(10), with thicknesses given in nm. This interfacial oxide layer between the TbCo and the Pd is key to properly fit the data due to the spin splitting around the $Q\approx0.05-0.07$ Å$^{-1}$ range which is most simply fit by introducing a magnetic component at the interface that is antiparallel to the net magnetization in the main layer. 

The inclusion of this layer in the model is physically justified by the growth process. While the sample remains under vacuum during the entire growth process, the change in base pressure between the two chambers combined with the reduced purity of the process gas is enough to create a small oxide layer on the TbCo surface. There is most likely a preferential oxidation of the Tb atoms which creates a layer that is magnetically Co-dominated instead of Tb-dominated. This kind of preferential oxidation has been studied previously through different means and it has been shown that even a small degree of oxidation can be enough to create a large shift in the compensation behaviour\cite{krupinski2021control, kiphart2025origin}. Despite this, the coupling between the two layers is most likely through the Co network with the Co-sublattices of each layer remaining coupled, similar to the effects previously seen in for instance GdCo/Pt systems where the interfacial moment is coupled to the transition-metal sublattice \cite{Swindells2020}. So, while the majority of the magnetization follows the applied magnetic field this small interfacial moment is antiparallel to the field.

Deuterium is loaded at three pressures; 5.8$\times10^{-2}$ mbar, 7.7$\times10^{-1}$ mbar and 10.7 mbar. At each pressure, reflectivity data is collected and fit in a constrained manner, using the fitting parameters from the sample under vacuum as the starting point. We then directly fit the thickness, nuclear SLD, and magnetic SLD of the TbCo layer. We recall that the SLD is related to the atomic concentration by,

\begin{equation}
    \rho_n = \frac{\rho N_{A} \sum_{i} c_i b_i}{\sum_{i} c_{i} M_{i}},
\end{equation}

where $\rho_\textnormal{n}$ is the nuclear SLD, $\rho$ is the mass density, $N_\textnormal{A}$ is Avogadro's number, $c_i$ is the atomic concentration of species $i$, and $M_i$ and $b_i$ are the corresponding atomic weights and coherent scattering lengths of the same element. Based on this, the deuterium concentration $C_\textnormal{D}$ in units of [D/M] can be obtained as \cite{Rehm1999}, 

\begin{equation}
    C_\textnormal{D} = [\frac{\rho_\textnormal{{M+D}} }{\rho_\textnormal{M} } \times \frac{t_\textnormal{{M+D}} }{t_\textnormal{M} } - 1] \times\frac{b_\textnormal{M} }{b_\textnormal{D} },
\end{equation}

where $\rho_M$ and $t_M$ are the initial SLD and thickness of the metal film obtained prior to deuterium loading, $\rho_{M+D}$ and $t_{M+D}$ are the SLD and thickness at a certain deuterium concentration, $b_D = 6.671$ fm is the coherent scattering length of $D$ and $b_M$ is the composition-weighted scattering length of the TbCo alloy. Obtained values of $C_\textnormal{D}$ are then converted into units of atomic percent. 

\begin{figure}
    \centering
    \includegraphics[width=\linewidth]{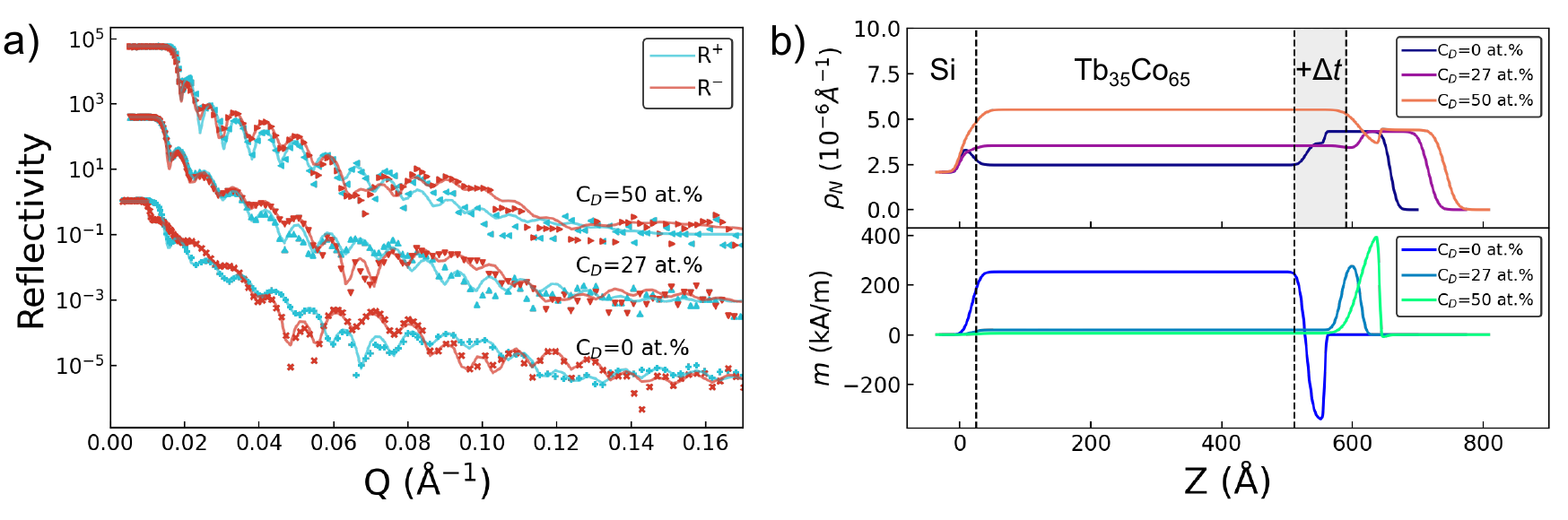}
    \caption{a) Change in polarized neutron reflectivity with D loading for three D concentrations. Data is offset for visual clarity. b) The corresponding nuclear SLD (top) and magnetization (bottom) profiles for each D concentration.}
    \label{fig:tb-rich_loading}
\end{figure}

The effect of deuterium loading is shown for three concentrations in Fig. \ref{fig:tb-rich_loading}. There are large changes in the features of the data relating to the changes in atomic composition and layer magnetization. These are seen most strongly around the critical edge, which is primarily sensitive to the TbCo layer, as it is the thickest layer: the critical edge moves to a higher point in $Q$ while the spin asymmetry of the two polarizations gradually reduces, corresponding to a loading of D and reduction in magnetic moment respectively seen in Fig. \ref{fig:tb-rich_loading}b).

We note that the oxide layer acts as a barrier to the migration of D between the Pd layer and the TbCo film. This slows down the diffusion of D which we take advantage of by measuring several scans sequentially at one D$_\textnormal{2}$ pressure. After the initial "rapid" loading of D at one pressure, the rate is slow enough to where this average scan can be reliably fit to one deuterium concentration. This is shown in figure \ref{fig:tb-rich_loading}a) where there is a clear change in the reflectivity data which is still possible to fit. As such we choose to show data for this sample in terms of the fitted concentration of D rather than the deuterium pressure.

\begin{figure}
    \centering
    \includegraphics[width=\linewidth]{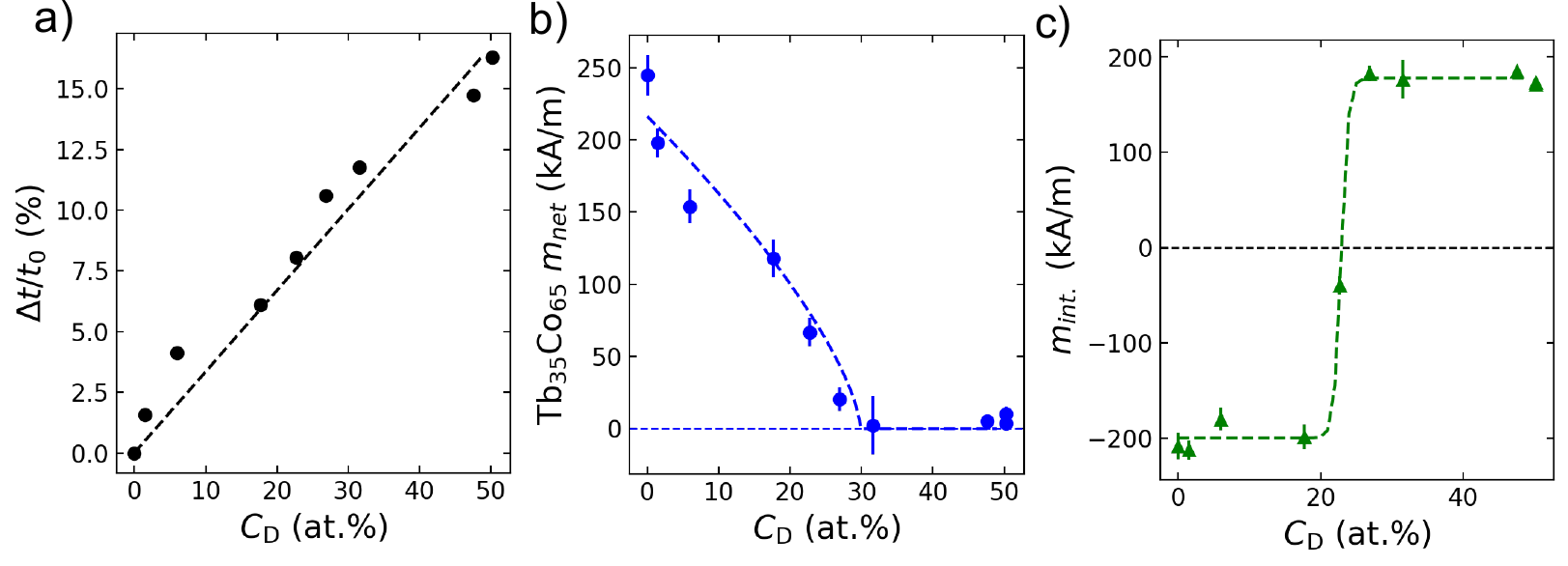}
    \caption{Fitting results extracted from the deuterium-loaded reflectometry profiles. a) The thickness expansion of the Tb$_{35}$Co$_{65}$ layer. b) The change in the net magnetic moment ($m_{\mathrm{net}}$) of the Tb$_{35}$Co$_{65}$ layer. c) Magnetic moment of the interfacial oxidized layer. Dashed lines are included as a guide to the eye.}
    \label{fig:tb-rich_summary}
\end{figure}

The results of the fitting are summarized in Fig. \ref{fig:tb-rich_summary}. We show the trends here of the three parameters that are most important for this study; the net magnetic moment $m_{\mathrm{net}}$ of the TbCo layer, the change in thickness, $t$, of the TbCo layer, and the magnetic moment of the interfacial layer, $m_\textnormal{int}$, between the TbCo and Pd.

The first point to notice is that there is a significant out-of-plane expansion of the film on hydrogen loading, achieving an expansion of 15\% at hydrogen concentrations of 50 at.\% D. This thickness expansion can be fit to a linear function which results in a relative thickness expansion,

\begin{equation}
    \frac{\Delta t}{t_0} = 0.335  C_\text{D},
\end{equation}

with $C_D$ expressed in units of at.\%. A linear expansion such as this has been observed in a variety of metallic films for hydrogen storage \cite{bylin2022hydrogen, Rehm1999}. In the metal-hydrogen literature, this coefficient is usually expressed using units of [D/M], and in this formalism the coefficient obtained is 0.18, which is in line with reported values for hydrogen storage materials \cite{Palsson2012}.




Moving on to the magnetic characterization, it is clear in the results of Fig. \ref{fig:tb-rich_summary}b) is that the introduction of deuterium alone causes a large initial drop in the magnetization followed by a smooth paramagnetic phase transition, which occurs at deuterium concentration of $C_D\approx28$ at.\%. Above this critical value of $C_D$ there is a small paramagnetic moment that aligns with the field, but the value of this is small and difficult to distinguish from fitting artifacts. 

An interesting additional result is seen in the change in the moment of the interfacial layer, $m_\textnormal{int}$. Prior to the phase transition, $m_\textnormal{int}$ remains opposed to the magnetic field direction but as soon as the magnetization $m_\textnormal{net}$ has reduced sufficiently this moment becomes unpinned and is free to align with the magnetic field. This supports the argument presented earlier that the interfacial moment $m_\textnormal{int}$ follows the Co-sublattice direction in the main TbCo layer. When the main TbCo layer becomes paramagnetic, the magnetization in the interfacial heavily-oxidized layer is able to freely re-align with the field, excluding any arguments of exchange bias or otherwise pinned magnetic moments. The concentration at which $m_\textnormal{int}$ is freely able to realign occurs at a value of $C_\textnormal{D}\approx23$ at.\%, which is lower than the paramagnetic transition which occurs at $C_D\approx28$ at.\%. The discrepancy between these two concentrations may indicate a weakening of the coupling between the two layers around the phase transition. 

\subsection{Influence of deuterium on out-of-plane magnetization}

We now discuss measurements performed on the Tb$_{14}$Co$_{86}$ sample. This sample is Co-rich and has PMA, meaning that the magnetization is pointing out-of-the-plane and perpendicular to the polarization vector of the neutron beam. In principle, this means that we should measure no magnetization in-plane and observe no spin splitting. However, we measure at a field of 300 mT to be able to more finely resolve the changes in magnetization such as if there is a sudden reorientation of the magnetization in-plane. This field is below the typical in-plane saturation fields for similar samples which can be of the order of $\approx 2$~T \cite{frisk2015tailoring}, but it is above the saturation field for samples that have in-plane magnetic anisotropy. A side-effect of applying this field is that there is a small rotation of the magnetization in-plane which leads to a measurable in-plane magnetic moment. 

\begin{figure}
    \centering
    \includegraphics[width=\linewidth]{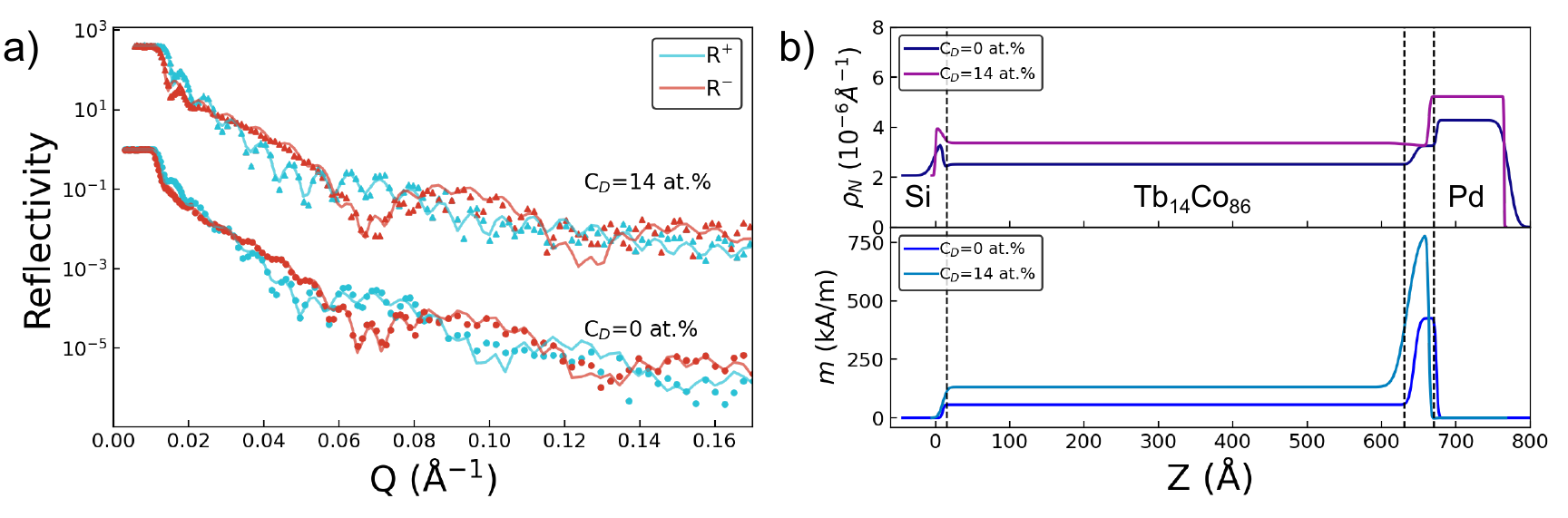}
    \caption{a) Change in neutron reflectivity with D loading for the Tb$_{14}$Co$_{86}$ sample under vacuum, and at a D$_\textnormal{2}$ pressure of 100 mBar (measured concentration of 14 at.\% D.) b) The corresponding nuclear and magnetic SLD profiles.}
    \label{fig:co-rich_loading}
\end{figure}

Fig. \ref{fig:co-rich_loading}a) shows both the initial reflectivity curves for the Co-rich sample. The sample is progressively loaded with increasing D$_\textnormal{2}$ pressure in the range of $6\times10^{-2} - 100$ mbar, and the final scan taken at a pressure of 100 mbar is also shown. As in the case of the Tb-rich sample, there is a change in the position of the critical edge which indicates a change in the nuclear scattering length density and that deuterium has been loaded into the sample. Similarly, we see a feature around $Q$=0.07 \AA $^{-1}$ that indicates the presence of an interfacial oxide layer between the TbCo and the Pd. However as compared with the Tb-rich sample the best fit is obtained when the magnetization of this layer is aligned parallel to the main TbCo layer - another indicator that the two layers are coupled via the Co sublattice.


\begin{figure}
    \centering
    \includegraphics[width=0.5\linewidth]{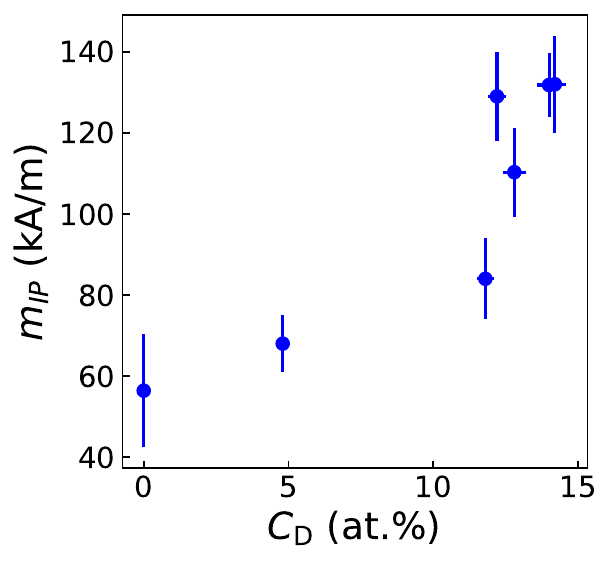}
    \caption{Fitted values of the in-plane net magnetization, $m_{\textnormal{IP}}$, for the Tb$_{14}$Co$_{86}$ sample. The magnetization initially increases linearly, then shows a sharp increase that could be the onset of a spin reorientation from out-of-plane to in-plane.}
    \label{fig:co-rich_summary}
\end{figure}

The fitted parameters are shown in Fig. \ref{fig:co-rich_summary}. Here we focus on the in-plane component, $m_\textnormal{IP}$, of the net magnetization and the achieved values of $C_\textnormal{D}$ as it was not possible to reliably observe a significant change in thickness with deuterium loading. At a temperature of 320 K and maximum D$_2$ pressure of 400 mbar, the maximum concentration of deuterium we were able to obtain was $C_\textnormal{D} = 14$ at.\%. This is likely due to the reduced hydrogen affinity of the alloy, since there are fewer Tb atoms, which makes it less thermodynamically favourable for deuterium ions to penetrate the oxide barrier between the Pd layer and TbCo film. In the absence of this oxide interface, the likelihood of loading higher concentrations of D would increase.

Correspondingly, we find that the best fit to the data incorporates a change in the SLD of the Pd layer which we attribute to deuterium accumulation in the layer. This is because the pressure used to load deuterium is much higher than for the Tb-rich sample and at these pressures the D-Pd system transitions to the solid solution $\alpha$ phase\cite{Lewis1982}. 

At these deuterium concentrations we see that the total in-plane moment at 300 mT increases by 130\%, with a sharp increase around the maximum deuterium concentration we were able to obtain which is indicative of a weakening out-of-plane anisotropy and the onset of a spin reorientation transition, though a full alignment was not observed. To examine the change in further detail and determine if there is any degree of reorientation we perform first quadrant measurements using polarized neutrons as a magnetometer. The instrument is aligned to the first spin asymmetry peak close to the critical edge, at $Q=0.0137$ Å$^{-1}$, and the reflectivity of both spin channels is measured as the magnetic field is varied. We first apply a magnetic field of -500 mT, then measure from 0 to 500 mT effectively performing one quadrant of a hysteresis loop. These fields are likely not enough to saturate the total film but are used to perform a minor hysteresis loop and to determine if there is some in-plane coercivity. 

\begin{figure}
    \centering
    \includegraphics[width=0.8\linewidth]{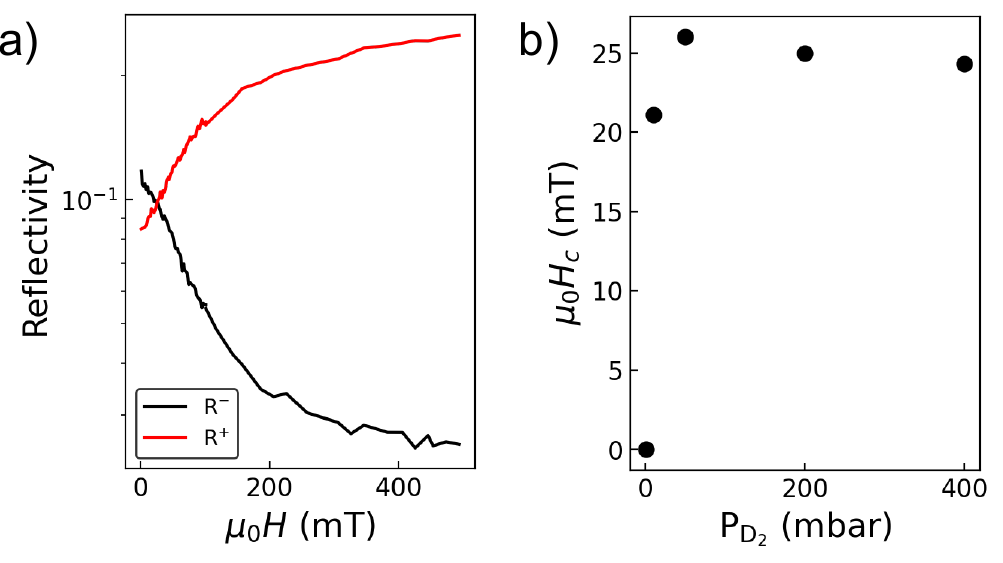}
    \caption{a) An example neutron magnetometry measurement taken at $Q=0.0137$ Å$^{-1}$, showing both the R$^+$ and R$^-$ reflectivities as the field is swept. b) The coercivities $H_c$ taken from these scans as the deuterium pressure is reduced.}
    \label{fig:neutron_magnetometry}
\end{figure}

An example of this measurement in shown in Fig. \ref{fig:neutron_magnetometry}a). The spin asymmetry is used as a probe of the magnetization state, which is zero at the coercive field when the R$^{+}$ and R$^{-}$ are equal. Fig. \ref{fig:neutron_magnetometry}b) shows the pressure-dependence of these measurements taken during the \textit{de-loading} of the sample, from high pressure to low pressure. The small in-plane coercivity that is measured at high pressures remains stable across a wide pressure range and very sharply reduces between 10 mbar and 1 mbar. 

This we interpret to show two things. First, that the increase in the in-plane magnetization is accompanied by a small coercivity in-plane where previously there was none, and second that the process is reversible and the PMA strength can be recovered by the de-loading of deuterium which here is achieved simply by reducing the D$_2$ gas pressure.

We cannot attribute these changes to thickness expansion as it was not possible to detect a significant increase in the thickness of the TbCo layer. Therefore, as compared with the Tb-rich sample, the loss of PMA may be more strongly related to more subtle changes such as the local atomic configuration that are not easily measureable as a change in thickness.

\section{Conclusions}

In conclusion, we have investigated the structural and magnetic changes in Tb$_x$Co$_{100-x}$ thin films with atmospheric deuterium loading, using samples with different Tb content\textemdash one that is Tb-rich and exhibits in-plane magnetic anisotropy, and one that is Co-rich and exhibits out-of-plane anisotropy. Using \textit{in-situ} polarized neutron reflectometry, we have measured the change in the in-plane magnetization profile whilst simultaneously measuring the deuterium concentration. We expect that these results seen for deuterium will be largely transferrable to hydrogen ions, which will be important for the field of hydrogen-based magneto-ionics.

For the in-plane Tb-rich sample, a clear dependence of the magnetization on deuterium concentration has been observed with a paramagnetic phase transition occurring at a concentration of $C_D\approx28$ at.\%. This reflectivity data is consistently fit very well to a single-layer model of the magnetization suggesting that the loading is uniform through the depth of the film. We attribute this magneto-ionic effect to the large increase in out-of-plane interatomic distances measured as a change in film thickness, which expands by $\approx10\%$ at this concentration. 

 In our Tb-rich sample, the Co sublattices of the TbCo layer and the oxidized interface layer remain coupled even as hydrogen is loaded and we were able to observe a hydrogen-dependence in the orientation of this interfacial layer purely by magneto-ionic manipulation of the TbCo layer. This sublattice-dependent coupling and subsequent magneto-ionic manipulation could be exploited in more complex heterostructures, particularly when the rare-earth moment is dominating as in our case.

For the out-of-plane Co-rich sample, we used the magnitude of the in-plane magnetic moment as an indirect measure of the out-of-plane anisotropy strength, with a larger moment indicating a reduction of this anisotropy and a greater degree of realignment in-plane. We saw an increase in in-plane magnetic moment corresponding to a reduction in out-of-plane anisotropy with deuterium loading, but were unable to observe a total reorientation of the moment in-plane. 

An important point from these results is the role of deuterium and hydrogen affinity in the loading processes of both of these films. While they both had a small oxide layer that acts as a barrier to loading, for the Tb-rich sample this only reduced the diffusion of deuterium whereas for the Co-rich sample this can completely inhibit significant loading. 

The difference is directly related to the rare-earth content of the material and, as seen in these results, has a large impact both in the robustness of the loading process and potentially in the mechanisms by which hydrogen can affect the magnetic properties. As such we would like to emphasize that the hydrogen affinity should always be considered when designing magneto-ionic devices based on rare-earth transition-metal alloys.


\section{Data Availability}

Data associated with the paper is publicly available at [TBD]. Data associated with the SuperADAM experiment are also available at doi:10.5291/ILL-DATA.5-72-53.

\section*{Acknowledgements}

The authors would like to acknowledge Kristina Komander and Eleni Ntemou for RBS measurements at the Tandem facility. M. P. G acknowledges support from the Swedish Research Council (Project No. 2023-06359). Accelerator operation at the Ion Technology Center national infrastructure is supported by the Swedish Research Council VR-RFI (grant \# 2019 00191). The SuperADAM
project acknowledges the Swedish Research Council for support. R. G. H. acknowledges the support of a Carl-Trygger Foundation Postdoctoral Fellowship (CTS 22:2039, grant holder G.A.)

\printbibliography

@article{bjorck2007genx,
  title={{GenX: an extensible X-ray reflectivity refinement program utilizing differential evolution}},
  author={Bj{\"o}rck, Matts and Andersson, Gabriella},
  journal={Journal of Applied Crystallography},
  volume={40},
  number={6},
  pages={1174--1178},
  year={2007},
  publisher={International Union of Crystallography},
  url = {https://doi.org/10.1107/S0021889807045086
}
}

@article{mayer1999simnra,
  title={{SIMNRA, a simulation program for the analysis of NRA, RBS and ERDA}},
  author={Mayer, Matej},
  journal={AIP conference proceedings},
  volume={475},
  pages={541--544},
  year={1999},
  organization={American Institute of Physics},
  url = {https://doi.org/10.1063/1.59188}
}

@article{de2022voltage,
  title={{Voltage control of magnetism with magneto-ionic approaches: Beyond voltage-driven oxygen ion migration}},
  author={de Rojas, Julius and Quintana, A and Rius, Gemma and Stefani, Christina and Domingo, Neus and Costa-Kr{\"a}mer, Jos{\'e} Luis and Men{\'e}ndez, Enric and Sort, Jordi},
  journal={Applied Physics Letters},
  volume={120},
  number={7},
  year={2022},
  publisher={AIP Publishing},
  url = {https://doi.org/10.1063/5.0079762}
}

@article{kossak2023voltage,
  title={{Voltage control of magnetic order in RKKY coupled multilayers}},
  author={Kossak, Alexander E and Huang, Mantao and Reddy, Pooja and Wolf, Daniel and Beach, Geoffrey S D},
  journal={Science Advances},
  volume={9},
  number={1},
  pages={eadd0548},
  year={2023},
  publisher={American Association for the Advancement of Science},
  url = {https://doi.org/10.1126/sciadv.add0548}
}

@article{ciuciulkaite2020magnetic,
  title={{Magnetic and all-optical switching properties of amorphous Tb$_x$ Co$_{100-x}$ alloys}},
  author={Ciuciulkaite, Agne and Mishra, Kshiti and Moro, Marcos V and Chioar, Ioan-Augustin and Rowan-Robinson, Richard M and Parchenko, Sergii and Kleibert, Armin and Lindgren, Bengt and Andersson, Gabriella and Davies, Carl S and others},
  journal={Physical Review Materials},
  volume={4},
  number={10},
  pages={104418},
  year={2020},
  publisher={American Physical Society},
  url = {https://doi.org/10.1103/PhysRevMaterials.4.104418}
}

@article{shuai2022local,
  title={{Local anisotropy control of Pt/Co/Ir thin film with perpendicular magnetic anisotropy by surface acoustic waves}},
  author={Shuai, Jintao and Ali, Mannan and Lopez-Diaz, Luis and Cunningham, John E and Moore, Thomas A},
  journal={Applied Physics Letters},
  volume={120},
  number={25},
  year={2022},
  publisher={AIP Publishing},
  url = {https://doi.org/10.1063/5.0097172}
}

@article{franke2015reversible,
  title={{Reversible electric-field-driven magnetic domain-wall motion}},
  author={Franke, K{\'e}vin J A and Van de Wiele, Ben and Shirahata, Yasuhiro and H{\"a}m{\"a}l{\"a}inen, Sampo J and Taniyama, Tomoyasu and van Dijken, Sebastiaan},
  journal={Phys. Rev. X},
  volume={5},
  number={1},
  pages={011010},
  year={2015},
  publisher={APS},
  url = {https://doi.org/10.1103/PhysRevX.5.011010}
}

@article{ederer2005weak,
  title={{Weak ferromagnetism and magnetoelectric coupling in bismuth ferrite}},
  author={Ederer, Claude and Spaldin, Nicola A},
  journal={Physical Review B—Condensed Matter and Materials Physics},
  volume={71},
  number={6},
  pages={060401},
  year={2005},
  publisher={APS},
  url = {https://doi.org/10.1103/PhysRevB.71.060401}
}

@article{bylin2022hydrogen,
  title={Hydrogen-induced volume changes, dipole tensor, and elastic hydrogen-hydrogen interaction in a metallic glass},
  author={Bylin, Johan and Malinovskis, Paulius and Devishvili, Anton and Scheicher, Ralph H and P{\'a}lsson, Gunnar K},
  journal={Physical Review B},
  volume={106},
  number={10},
  pages={104110},
  year={2022},
  publisher={APS},
  url = {https://doi.org/10.1103/PhysRevB.106.104110}
}

@article{krupinski2021control,
  title={{Control of magnetic properties in ferrimagnetic GdFe and TbFe thin films by He$^+$ and Ne$^+$ irradiation}},
  author={Krupinski, Michal and Hintermayr, Julian and Sobieszczyk, Pawel and Albrecht, Manfred},
  journal={Physical Review Materials},
  volume={5},
  number={2},
  pages={024405},
  year={2021},
  publisher={APS},
  url = {https://doi.org/10.1103/PhysRevMaterials.5.024405}
}

@article{bauer2015magneto,
  title={Magneto-ionic control of interfacial magnetism},
  author={Bauer, Uwe and Yao, Lide and Tan, Aik Jun and Agrawal, Parnika and Emori, Satoru and Tuller, Harry L and Van Dijken, Sebastiaan and Beach, Geoffrey S D},
  journal={Nature materials},
  volume={14},
  number={2},
  pages={174--181},
  year={2015},
  publisher={Nature Publishing Group UK London}
}

@article{hunt2025control,
  title={{Control of ferrimagnetic compensation and perpendicular anisotropy in Tb$_x$Co$_{(100- x)}$ with H$^+$ ion implantation}},
  author={Hunt, Robbie G and Moldarev, Dmitrii and Grassi, Mat{\'\i}as P and Primetzhofer, Daniel and Andersson, Gabriella},
  journal={Physical Review Materials},
  volume={9},
  number={3},
  pages={034409},
  year={2025},
  publisher={APS}
}

@article{zamani2013,
title = {{Tuning magnetic properties by hydrogen implantation in amorphous Fe$_{100-x}$Zr$_x$ thin films}},
journal = {Journal of Magnetism and Magnetic Materials},
volume = {346},
pages = {138-141},
year = {2013},
issn = {0304-8853},
doi = {https://doi.org/10.1016/j.jmmm.2013.07.031},
url = {https://www.sciencedirect.com/science/article/pii/S0304885313005167},
author = {Atieh Zamani and Anders Hallén and Per Nordblad and Gabriella Andersson and Björgvin Hjörvarsson and Petra E. Jönsson},
}

@article{Swindells2020,
  title = {{Proximity-induced magnetism in Pt layered with rare-earth–transition-metal ferrimagnetic alloys}},
  volume = {2},
  ISSN = {2643-1564},
  url = {http://dx.doi.org/10.1103/PhysRevResearch.2.033280},
  %DOI = {10.1103/physrevresearch.2.033280},
  number = {3},
  journal = {Physical Review Research},
  publisher = {American Physical Society (APS)},
  author = {Swindells,  C. and Nicholson,  B. and Inyang,  O. and Choi,  Y. and Hase,  T. and Atkinson,  D.},
  year = {2020},
  month = aug 
}

@article{Rehm1999,
  title = {Hydrogen concentration and its relation to interplanar spacing and layer thickness of 1000-Å Nb(110) films duringin situhydrogen charging experiments},
  volume = {59},
  ISSN = {1095-3795},
  url = {http://dx.doi.org/10.1103/PhysRevB.59.3142},
  DOI = {10.1103/physrevb.59.3142},
  number = {4},
  journal = {Physical Review B},
  publisher = {American Physical Society (APS)},
  author = {Rehm,  Ch. and Fritzsche,  H. and Maletta,  H. and Klose,  F.},
  year = {1999},
  month = jan,
  pages = {3142–3152}
}

@Article{telegrin2022,
AUTHOR = {Telegin, Andrei and Sukhorukov, Yurii},
TITLE = {Magnetic Semiconductors as Materials for Spintronics},
JOURNAL = {Magnetochemistry},
VOLUME = {8},
YEAR = {2022},
NUMBER = {12},
ARTICLE-NUMBER = {173},
URL = {https://www.mdpi.com/2312-7481/8/12/173},
ISSN = {2312-7481},
%DOI = {10.3390/magnetochemistry8120173}
}

@article{Mostovoy2024,
  title = {Multiferroics: different routes to magnetoelectric coupling},
  volume = {2},
  ISSN = {2948-2119},
  url = {http://dx.doi.org/10.1038/s44306-024-00021-8},
 % DOI = {10.1038/s44306-024-00021-8},
  number = {1},
  journal = {npj Spintronics},
  publisher = {Springer Science and Business Media LLC},
  author = {Mostovoy,  Maxim},
  year = {2024},
  month = jun 
}

@article{gossler2021nanoporous,
  title={Nanoporous Pd1- xCox for hydrogen-intercalation magneto-ionics},
  author={G{\"o}{\ss}ler, Markus and Topolovec, Stefan and Krenn, Heinz and W{\"u}rschum, Roland},
  journal={APL Materials},
  volume={9},
  number={4},
  year={2021},
  publisher={AIP Publishing}
}

@article{bischoff2024magneto,
  title={Magneto-Ionic Control of Coercivity and Domain-Wall Velocity in Co/Pd Multilayers by Electrochemical Hydrogen Loading},
  author={Bischoff, Madeleine and Ehrler, Rico and Engelhardt, Felix and Hellwig, Olav and Leistner, Karin and G{\"o}{\ss}ler, Markus},
  journal={Advanced Functional Materials},
  volume={34},
  number={40},
  pages={2405323},
  year={2024},
  publisher={Wiley Online Library}
}

@article{das2018detection,
  title={Detection of hydrogen by the extraordinary Hall effect in CoPd alloys},
  author={Das, SS and Kopnov, G and Gerber, A},
  journal={Journal of Applied Physics},
  volume={124},
  number={10},
  year={2018},
  publisher={AIP Publishing}
}

@article{klose1997continuous,
  title={Continuous and reversible change of the magnetic coupling in an Fe/Nb multilayer induced by hydrogen charging},
  author={Klose, F and Rehm, Ch and Nagengast, D and Maletta, H and Weidinger, A},
  journal={Physical review letters},
  volume={78},
  number={6},
  pages={1150},
  year={1997},
  publisher={APS}
}

@article{labergerie2001hydrogen,
  title={Hydrogen induced change of the atomic magnetic moments in Fe/V-superlattices},
  author={Labergerie, D and Westerholt, K and Zabel, H and Hj{\"o}rvarsson, Bj{\"o}rgvin},
  journal={Journal of magnetism and magnetic materials},
  volume={225},
  number={3},
  pages={373--380},
  year={2001},
  publisher={Elsevier}
}

@article{cousin2020introduction,
  title={An introduction to neutron reflectometry},
  author={Cousin, Fabrice and Fadda, Giulia},
  journal={EPJ web of conferences},
  volume={236},
  pages={04001},
  year={2020},
  organization={EDP Sciences},
}

@article{devishvili2013superadam,
  title={SuperADAM: Upgraded polarized neutron reflectometer at the Institut Laue-Langevin},
  author={Devishvili, A and Zhernenkov, K and Dennison, Andrew JC and Toperverg, BP and Wolff, Max and Hj{\"o}rvarsson, Bj{\"o}rgvin and Zabel, H},
  journal={Review of Scientific Instruments},
  volume={84},
  number={2},
  year={2013},
  publisher={AIP Publishing}
}

@article{vorobiev2015recent,
  title={Recent upgrade of the polarized neutron reflectometer Super ADAM},
  author={Vorobiev, Alexei and Devishvilli, Anton and Palsson, Gunnar and Rundl{\"o}f, H{\aa}kan and Johansson, Niklas and Olsson, Anders and Dennison, Andrew and Wollf, Max and Giroud, Benjamin and Aguettaz, Olivier and others},
  journal={Neutron News},
  volume={26},
  number={3},
  pages={25--26},
  year={2015},
  publisher={Taylor \& Francis}
}

@article{frisk2015tailoring,
  title={Tailoring anisotropy and domain structure in amorphous TbCo thin films through combinatorial methods},
  author={Frisk, Andreas and Magnus, Fridrik and George, Sebastian and Arnalds, Unnar B and Andersson, Gabriella},
  journal={Journal of Physics D: Applied Physics},
  volume={49},
  number={3},
  pages={035005},
  year={2015},
  publisher={IOP Publishing}
}

@article{kiphart2025origin,
  title={Origin of ion bombardment induced Tb oxidation in Tb/Co multilayers},
  author={Kiphart, Daniel and Krupi{\'n}ski, M and Mitura-Nowak, Marzena and Micha{\l}owski, PP and Kowacz, Mateusz and Schmidt, Marek and Stobiecki, Feliks and Chaves-O’Flynn, Gabriel David and Ku{\'s}wik, Piotr},
  journal={Applied Surface Science},
  volume={685},
  pages={162090},
  year={2025},
  publisher={Elsevier}
}

@article{hjorvarsson1997reversible,
  title={Reversible tuning of the magnetic exchange coupling in Fe/V (001) superlattices using hydrogen},
  author={Hj{\"o}rvarsson, B and Dura, JA and Isberg, P and Watanabe, T and Udovic, TJ and Andersson, G and Majkrzak, CF},
  journal={Physical review letters},
  volume={79},
  number={5},
  pages={901},
  year={1997},
  publisher={APS}
}

@book{alefeld1978hydrogen2,
  title = {Hydrogen in Metals II},
  ISBN = {9783540358015},
  ISSN = {1437-0859},
  url = {http://dx.doi.org/10.1007/3-540-08883-0},
  DOI = {10.1007/3-540-08883-0},
  journal = {Topics in Applied Physics},
  publisher = {Springer Berlin Heidelberg},
  year = {1978}
}

@book{alefeld1978hydrogen1,
  title = {Hydrogen in Metals I},
  ISBN = {9783540358923},
  ISSN = {1437-0859},
  url = {http://dx.doi.org/10.1007/3-540-08705-2},
  DOI = {10.1007/3-540-08705-2},
  journal = {Topics in Applied Physics},
  publisher = {Springer Berlin Heidelberg},
  year = {1978}
}

@article{baibich1988GMR,
  title = {Giant Magnetoresistance of (001)Fe/(001)Cr Magnetic Superlattices},
  author = {Baibich, M. N. and Broto, J. M. and Fert, A. and Van Dau, F. Nguyen and Petroff, F. and Etienne, P. and Creuzet, G. and Friederich, A. and Chazelas, J.},
  journal = {Phys. Rev. Lett.},
  volume = {61},
  issue = {21},
  pages = {2472--2475},
  numpages = {0},
  year = {1988},
  publisher = {American Physical Society},
  doi = {10.1103/PhysRevLett.61.2472},
  url = {https://link.aps.org/doi/10.1103/PhysRevLett.61.2472}
}

@article{bowen2001large,
  title={Large magnetoresistance in Fe/MgO/FeCo (001) epitaxial tunnel junctions on GaAs (001)},
  author={Bowen, M and Cros, Vicent and Petroff, F and Fert, Albert and Mart{\i}nez Boubeta, C and Costa-Kr{\"a}mer, Jos{\'e} Luis and Anguita, Jos{\'e} Virgilio and Cebollada, Alfonso and Briones, F and De Teresa, JM and others},
  journal={Applied Physics Letters},
  volume={79},
  number={11},
  pages={1655--1657},
  year={2001},
  publisher={American Institute of Physics}
}

@article{dee2008magnetic,
  title={Magnetic tape for data storage: An enduring technology},
  author={Dee, Richard H},
  journal={Proceedings of the IEEE},
  volume={96},
  number={11},
  pages={1775--1785},
  year={2008},
  publisher={IEEE}
}

@article{ma2025magneto,
  title={Magneto-Ionic Engineering of Antiferromagnetically RKKY-Coupled Multilayers},
  author={Ma, Zheng and Arredondo-L{\'o}pez, Aitor and Wrona, Jerzy and Herrero-Mart{\'\i}n, Javier and Langer, Juergen and Berthold, Ocker and Pellicer, Eva and Men{\'e}ndez, Enric and Sort, Jordi},
  journal={Advanced Materials},
  volume={37},
  number={19},
  pages={2415393},
  year={2025},
  publisher={Wiley Online Library}
}

@article{kutuzau2025additive,
  title={Additive-enhanced hydrogen-and hydroxide-based magneto-ionic control in Ni films},
  author={Kutuzau, Maksim and G{\"o}{\ss}ler, Markus and Topolovec, Stefan and Schiemenz, Sandra and Wolf, Daniel and Richter, Manuel and Nielsch, Kornelius and Leistner, Karin},
  journal={Physical Review Materials},
  volume={9},
  number={11},
  pages={114408},
  year={2025},
  publisher={APS}
}

@article{lopez2024room,
  title={Room-Temperature Solid-State Nitrogen-Based Magneto-Ionics in Co$_x$Mn$_{1- x}$N Films},
  author={L{\'o}pez-Pint{\'o}, Nicolau and Jensen, Christopher J and Chen, Zhijie and Tan, Zhengwei and Ma, Zheng and Liedke, Maciej Oskar and Butterling, Maik and Wagner, Andreas and Herrero-Mart{\'\i}n, Javier and Men{\'e}ndez, Enric and others},
  journal={Advanced Functional Materials},
  volume={34},
  number={42},
  pages={2404487},
  year={2024},
  publisher={Wiley Online Library}
}

@article{yan2015electrical,
  title={Electrical control of Co/Ni magnetism adjacent to gate oxides with low oxygen ion mobility},
  author={Yan, YN and Zhou, XJ and Li, F and Cui, B and Wang, YY and Wang, GY and Pan, F and Song, C},
  journal={Applied Physics Letters},
  volume={107},
  number={12},
  year={2015},
  publisher={AIP Publishing}
}

@ARTICLE{Lewis1982,
  title     = "The palladium-hydrogen system",
  author    = "Lewis, F A",
  journal   = "Platinum Metals Review",
  publisher = "Johnson Matthey",
  volume    =  26,
  number    =  1,
  pages     = "20--27",
  month     =  jan,
  year      =  1982,
  language  = "en"
}

@article{Palsson2012,
  title = {Hydrogen site occupancy and strength of forces in nanosized metal hydrides},
  volume = {85},
  ISSN = {1550-235X},
  url = {http://dx.doi.org/10.1103/PhysRevB.85.195407},
%  DOI = {10.1103/physrevb.85.195407},
  number = {19},
  journal = {Physical Review B},
  publisher = {American Physical Society (APS)},
  author = {Pálsson,  Gunnar K. and W\"{a}lde,  Moritz and Amft,  Martin and Wu,  Yuanyuan and Ahlberg,  Martina and Wolff,  Max and Pundt,  Astrid and Hj\"{o}rvarsson,  Bj\"{o}rgvin},
  year = {2012},
  month = may 
}

\end{document}